\documentstyle[12pt,epsfig]{article}
\def\beqra{\begin{eqnarray}} 
\def\eeqra{\end{eqnarray}}
\def\beq{\begin{equation}}      
\def\eeq{\end{equation}}
\def\ds{\displaystyle}

\def\Vb{\bar{V}}
\def\phb{\bar{\phi}}
\def\rhb{\bar{\rho}}
\def\L{\Lambda}

\def \lta {\mathrel{\vcenter
     {\hbox{$<$}\nointerlineskip\hbox{$\sim$}}}}

{

\newcommand{\Tr}{\mathop{\mathrm{Tr}}}

\def\i{i}

\begin{document}
\sloppy
%
%
\begin{titlepage}
\begin{flushright}
CERN-TH/96-341\\
\end{flushright}
\vspace{3cm}
\centerline{\Large\bf Inverse Symmetry Breaking 
}
\vspace{10pt}
\centerline{\Large\bf 
and the Exact Renormalization Group}
\vspace{24pt}
\centerline{\large M. Pietroni, N. Rius\footnote{Permanent address
after 1/11/96: Departament de F\'{\i}sica Te\`orica, 
Universitat de Val\`encia and IFIC, Val\`encia, Spain.}
and N. Tetradis}
\vspace{10pt}
\centerline{\it Theory Division, CERN}
\centerline{\it CH-1211 Geneva 23, Switzerland}

\vspace{4cm}
\begin{abstract}
We discuss the question of inverse symmetry breaking at 
non-zero temperature using the exact renormalization group.
We study a two-scalar theory and concentrate on the 
nature of the  
phase transition during which the symmetry is broken. 
We also examine the 
persistence of symmetry breaking at temperatures higher than the
critical one. 
\end{abstract}
\vspace{4cm}
CERN-TH/96-341\\
November 1996
\end{titlepage}

It has been known for a long time \cite{Weinberg,Mohapatra,Langacker}, 
that simple  multiscalar models can exhibit
an anti-intuitive behaviour associated with more broken symmetry as the 
temperature is increased. We refer to this behaviour as 
inverse symmetry breaking. 
This possibility may have remarkable consequences for cosmology, solving 
the problem of  topological defects, thanks to the fact that 
the phase transition leading to their formation may never 
have  occurred during 
the thermal history of the Universe \cite{Dvali}.
The existence of the 
phenomenon for supersymmetric theories has been discussed in
refs. \cite{Tamvakis}. However, doubts have been raised 
on the validity of these results, which are based on the
one-loop approximation to the non-zero-temperature effective potential. 
This approximation is known to be unreliable for the discussion
of many aspects of phase transitions.
Recently the effect of next-to-leading-order contributions within
perturbation theory has been investigated in ref. 
\cite{Bimonte1}. This has been done through
the study of gap equations, which are
equivalent to a resummation of the super-daisy diagrams of the
perturbative series. Large subleading corrections have been identified, 
which lead to a sizeable reduction of the parameter space where inverse 
symmetry breaking occurs. The question of inverse symmetry breaking
has also been studied through the use of the renormalization group,
with similar conclusions \cite{Roos}.  
A variational approach has been employed in ref. \cite{Camelia}. 
Contrary to the results of the above studies, 
a large-$N$ analysis seems to indicate
that symmetry is always restored at high temperature  \cite{Fujimoto}.
However, the validity of this claim has recently been questioned in ref. 
\cite{Orloff}. A finite-lattice calculation also supports the symmetry
restoration at sufficiently high temperature \cite{Bimonte2}, even though 
the relevance of this result for the continuum limit is not clear.

In this letter we study the question of inverse symmetry breaking
by employing the exact renormalization group
\cite{Wilson}. Our formalism is 
similar to that of ref. \cite{Roos}. Our approach, however, is based on 
the real-time formulation of high-temperature field theories. 
We investigate a two-scalar theory and we identify the universal 
behaviour associated with the symmetry-breaking 
phase transition. 
Our study is based on an evolution 
equation for the potential of the non-zero temperature theory. 
Using a polynomial ansatz for the potential,
we solve this equation 
and verify the
conclusion of ref. \cite{Roos} that inverse symmetry breaking is 
confirmed by the renormalization-group approach. 
Moreover, 
we explore the parameter space that
leads to inverse symmetry breaking and compare it with the perturbative
predictions. 
In order to study the phase transition, we
go beyond the polynomial ansatz and consider a general dependence of the
potential on the field that develops an expectation value when
the symmetry is broken. The resulting partial differential equation
is solved numerically with the use of appropriate algorithms
\cite{Adams}. The Wilson-Fisher fixed point of the effective
three-dimensional theory is shown to govern the dynamics near the
second-order phase transition. 

We consider the simplest model that exhibits inverse symmetry breaking:
a two-scalar model with $Z_2 \times Z_2$ symmetry. 
The tree-level potential is given by 
\beq
V_{tr}(\phi_1, \phi_2) = \frac{1}{2} m_1^2 \phi_1^2 +\frac{1}{2} m_2^2 \phi_2^2
+ \frac{1}{4} \lambda_1 \phi_1^4 + \frac{1}{4} \lambda_2 \phi_2^4 -
\frac{1}{2}\lambda_{12} \phi_1^2 \phi_2^2~.
\label{pot}
\eeq
This potential is bounded for $\lambda_{1,2}>0$ 
and 
\beq
\lambda_1 \lambda_2 > \lambda_{12}^2~.
\label{stab} \eeq
The thermal correction to the above potential at the one-loop level 
is given by the well-known expression \cite{Dolan} 
\beq
\Delta V_{T}(\phi_1, \phi_2) = T \int_0^{\infty} 
\frac{d k}{2 \pi^2} k^2 \Tr\log
\left[1-\exp\left(-\frac{1}{T}\sqrt{k^2+{\cal M}_{tr}^2}\right)\right]~.
\label{dvt}
\eeq
Here ${\cal M}_{tr}^2$ indicates the matrix of second derivatives of the 
tree-level effective potential
\beq
\left[{\cal M}_{tr}^2(\phi_1, \phi_2)\right]_{i,j} = 
\frac{\partial^2 V_{tr}(\phi_1, \phi_2)}{\partial \phi_i \partial \phi_j}~, 
\;\;\;\;\;\;\;\;i,j = 1,2 ~.
\eeq
When both eigenvalues of ${\cal M}_{tr}^2$ are much smaller 
than $T^2$ (which happens for sufficiently small couplings), 
the leading field-dependent 
correction takes the form 
\beq
\ds \Delta V_{T}(\phi_1, \phi_2) \simeq 
\ds
\frac{T^2}{24}\left[(3\lambda_1 - 
\lambda_{12}) \phi_1^2 + (3 \lambda_2 - \lambda_{12}) \phi_2^2 \right]~+~\dots
\label{pert}
\eeq
For the parameter range 
\beq
3 \lambda_1 - \lambda_{12} < 0~,
\label{pertpred}
\eeq
which can be consistent with the stability condition of eq. (\ref{stab}), 
the thermal correction for the mass term of the $\phi_1$ field is negative. 
If the system is in the symmetric phase at zero temperature with
$m_{1,2}^2>0$, there will be a critical temperature 
$T_{cr}^2 = 12 m_1^2/(\lambda_{12}
-3\lambda_1)$ above which the symmetry will be broken
\footnote{
For $3 \lambda_2 - \lambda_{12} < 0~$ the symmetry is broken in the
$\phi_2$ direction. 
The discussion of this case 
is completely analogous to the one we consider. Notice that 
it is not possible to break the symmetry in both the 
$\phi_1$ and $\phi_2$ directions, because of the stability condition
of eq. (\ref{stab}).}.
If the system is 
in the broken phase at $T=0$, the symmetry will never be 
restored by thermal corrections. 

Our aim is to discuss the above scenario
in the context of the Wilson approach to the renormalization group. 
The main ingredient in this approach is an exact flow equation that 
describes how the effective action of the system evolves
as the ultraviolet cutoff is lowered. We consider the lowest order in a
derivative expansion of the effective action, which contains 
a general effective potential and a standard kinetic term.
At non-zero temperature this approach can be formulated either in the 
imaginary-time \cite{Nick1} or in the real-time formalism \cite{Max}. 
In the latter formulation, the evolution of the potential lowering 
the cutoff scale $\Lambda$ is  given by 
the partial differential equation
\cite{Max}
\beq
\L\frac{\partial\:}{\partial \L} 
V_{\L}(\phi_1, \phi_2) = -T\frac{\L^3}{2 \pi^2}
\Tr\left\{ \log\left[1-\exp\left(-\frac{1}{T}\sqrt{\L^2+{\cal M}_\L^2}\right)
\right]\right\}~,
\label{4dr}
\eeq
where 
\beq
\left[{\cal M}_\L^2(\phi_1, \phi_2)\right]_{i,j} = 
\frac{\partial^2 V_{\L}(\phi_1, \phi_2)}{\partial \phi_i \partial \phi_j}~, 
\;\;\;\;\;\;\;\;i,j = 1,2 ~.
\label{massma}
\eeq
Notice the formal similarity with the Dolan-Jackiw one-loop result of eq. 
(\ref{dvt}). The main difference is that 
the `running' mass matrix 
in the exponential of eq. (\ref{massma}) replaces the  
tree-level one in eq. (\ref{dvt}). 
The initial condition for the above equation, at 
a scale $\L_0 \gg T$, is the renormalized effective
potential at zero temperature.
In this work 
we consider small quartic couplings,  so that the  
logarithmic corrections of the zero-temperature theory can be
safely neglected. The initial condition for the evolution is a zero-temperature
potential given by eq. (\ref{pot}).  
Integrating  
the evolution equation (\ref{4dr}), we obtain the non-zero-temperature
effective potential
in the limit $\L\rightarrow 0$.
In the approximation that the `running' mass matrix  
on the r.h.s. of the evolution equation
is taken to be constant, equal to 
the tree-level mass matrix, the integration 
reproduces the perturbative result of eq. (\ref{dvt}).
The non-trivial behaviour that we describe in the
following paragraphs is obtained when the 
full scale-dependence of the mass matrix is taken into account. 
An interesting limit corresponds to the evolution at scales 
$\L \ll T$. Let us denote by $\left[
\tilde{\cal M}_{\L}^2 \right]_l$ the eigenvalues of the
mass matrix that satisfy $\left[ \tilde{\cal M}_{\L}^2 \right]_l \lta \L^2$. 
The remaining eigenvalues
correspond to decoupled 
massive modes that do not contribute to the evolution at 
the scale $\L$ \cite{Nick2}.
Keeping the leading contribution in the r.h.s. of 
eq. (\ref{4dr}) and omitting the field-independent terms, we obtain 
\beq  
\L\frac{\partial\:}{\partial \L} \left( \frac{V_\L(\phi_1, \phi_2)}{T} 
\right) =
-\frac{\L^3}{4\pi^2} \Tr \left\{ \log\left(\L^2 + 
\left[ \tilde{\cal M}_{\L}^2 \right]_l
\right)\right\}~.
\label{3dr}
\eeq
The rescaled potential
$V_\L/T$ has dimensions (mass$)^3$ and its evolution is 
typical of that of 
a three-dimensional theory
\cite{Nick2,Morris}. 
In the limit $\L \ll T$, 
dimensional reduction takes place, and a four-dimensional 
theory at non-zero temperature behaves as an effective three-dimensional 
one at $T=0$. 
We can cast eq. (\ref{3dr}) in a form that does not depend explicitly
on the scale $\L$ by defining the dimensionless parameters:
\beq
\phb_{1,2} = \frac{\phi_{1,2}}{\sqrt{\L T}}~,~~~~~~~~~~~
\Vb_\L(\phb_1, \phb_2) = 
\frac{V_\L(\phi_1, \phi_2)}{\L^3 T}~,~~~~~~~~~~~
\left[ \bar{\cal M}_{\L}^2 \right]_l = 
\frac{\left[ \tilde{\cal M}_{\L}^2\right]_l}{\L^2}~.
\label{resc} \eeq 
The dimensionless mass matrix 
$\bar{\cal M}_{\L}^2$ is related to the rescaled potential 
$\Vb$ through an equation analogous to eq. (\ref{massma}).
The evolution equation now reads
\beq
\L\frac{\partial\:}{\partial \L} \Vb_\L(\phb_1, \phb_2)
= -3 \Vb
-\frac{1}{4\pi^2} \Tr \left\{
\log\left(1 + \left[ \bar{\cal M}_{\L}^2 \right]_l
\right)\right\}~,
\label{3drr}
\eeq
where again we have omitted the field-independent terms. 
The scale-invariant (fixed-point) 
solutions of the effective three-dimensional theory 
can be obtained from the above equation for 
$\L \partial \Vb_\L  / \partial \L =0$.

Finding the solution of eq. (\ref{4dr}) is a difficult
task. An approximate solution can be obtained \cite{Nick1,Roos}
by expanding the potential in a power series in the fields. 
In this way
the partial differential equation (\ref{4dr}) is transformed 
into an infinite
system of ordinary differential equations for the 
coefficients of the expansion. 
This system can be solved approximately by truncation at a finite 
number of equations. 
In effect,  
the potential is approximated by a finite-order polynomial. 
As a first step, we follow this procedure and define   
the running masses and couplings at the origin 
\beq
m^2_{1,2}(\L)= \left. \frac{\partial^2 V_\L}{\partial \phi_{1, 2}^2}
\right|_{\phi_{1,2}=0},\:\:\:\:\:\:\:
\lambda_{1,2}(\L)= \left. \frac{1}{6}
\frac{\partial^4 V_\L}{\partial \phi_{1, 2}^4}
\right|_{\phi_{1,2}=0},
\:\:\:\:\:\:
\lambda_{12}(\L)= - \left. 
\frac{1}{2}\frac{\partial^4 V_\L}{\partial \phi_{1}^2
\partial \phi_{2}^2}
\right|_{\phi_{1,2}=0}~.
\label{param}
\eeq
The corresponding evolution equations can be obtained by 
differentiating eq. (\ref{4dr}) and neglecting the higher derivatives
of the potential. We find 
\beqra\label{trunc}
\ds \L \frac{\partial \:}{\partial\L} m_{1,2}^2 &=&\ds - 6 C_{1,2} 
\lambda_{1,2} 
+ 2 C_{2,1} \lambda_{12} \nonumber \\
\ds \L \frac{\partial \:}{\partial\L} \lambda_{1,2} &=&\ds - 18 D_{1,2}
 \lambda_{1,2}^2 
- 2 D_{2,1} \lambda_{12}^2  \\
\ds \L \frac{\partial \:}{\partial\L} \lambda_{12} &=&\ds - 6 D_1
 \lambda_{1} \lambda_{12}  - 6 D_2 \lambda_{2} \lambda_{12}
  +8 \frac{C_1 - C_2}{{m_1}^2 - m_{2}^2} \lambda_{12}^2~,\nonumber
\eeqra
with
\beq
C_{1, 2}=\frac{\L^3}{4 \pi^2}\frac{N(\omega_{1,2})}{\omega_{1,2}}~, 
\:\:\:\:\:\:\:\:
D_{1, 2} = \frac{\partial 
C_{1,2}}{\partial m_{1,2}^2}~,\:\:\:\:\:\:\:\: 
\omega_{1,2}^2=\L^2 + m_{1,2}^2~,
\label{def1} \eeq
and $N(\omega)=\left[ 
\exp(\omega/T)-1 \right]^{-1}$ the Bose-Einstein distribution 
function.
For $\omega_{1,2} \ll T$ 
we have
\beq
C_{1,2}\rightarrow \frac{\L^3}{4\pi^2}\frac{T}{\L^2+m_{1,2}^2}~,
\label{limit} \eeq
and the above equations agree with those considered in ref. \cite{Roos}
in the same limit.
For $\omega_{1,2} \gg T$ there is no 
running, because of 
the exponential suppression in the Bose-Einstein function. 

We have solved numerically the system of equations (\ref{trunc}) 
and determined the range of zero-temperature parameters
that lead to inverse symmetry breaking. 
In fig.~1 we present the results for a zero-temperature theory
with positive mass terms $m^2_1(\L_0)=m_2^2(\L_0)$ and $\lambda_2(\L_0)=0.3$. 
The temperature has been chosen much higher than the 
critical one (\mbox{$T=500 m_1(\L_0)$}).  
The system  (\ref{trunc}) has been integrated from $\Lambda_0 \gg T$ down to
\mbox{$\L = 0$}, where the thermally corrected masses and 
couplings at non-zero temperature have been obtained. 
A negative value for the mass term $m^2_1$ at $\L=0$
has been considered as 
the signal of inverse symmetry breaking. This has been achieved 
for the parameter range of $\lambda_1(\L_0)$, $\lambda_{12}(\L_0)$ 
above the line (a) in fig.~1. In the same 
figure we plot the stability bound of  
eq. (\ref{stab}). The allowed range is below the line (b). 
We also include the perturbative prediction 
of eq. (\ref{pertpred}) for the range that leads to inverse symmetry
breaking. It lies above the line (c).  
The phenomenon of inverse symmetry breaking is confirmed by our study,
in agreement with ref. \cite{Roos}, where the imaginary-time 
formulation of the renormalization-group approach has been used. 
We observe that the renormalization-group treatment eliminates 
a large part of the parameter space allowed by perturbative theory.
This is in agreement with the results of ref. \cite{Bimonte1}, where
the gap-equation approach has been followed.
Notice that lines (a) and (c) approach each other near the origin, where
perturbation theory becomes more reliable. 

The reliability of our conclusions crucially depends  on whether 
the solution of the system of truncated equations 
(\ref{trunc}) provides an approximate solution to the 
full partial differential equation (\ref{4dr}). 
In ref. \cite{Roos}, this has been checked by increasing the
level of truncations and verifying the convergence of the results. 
We follow here a different approach that relies on the 
numerical integration of eq. (\ref{4dr}) through the algorithms 
discussed in ref.~\cite{Adams}. These algorithms have been used 
for the integration of the evolution equations for potentials
that depend on one field only \cite{Adams,Nick3}. 
The generalization to the 
two-field case is straightforward. However, limitations in 
computer time have prevented us from reaching full numerical stability for
the results.
For this reason, we restrict our discussion of eq. (\ref{4dr})
along the $\phi_1$ axis, which is
the direction of expected symmetry breaking for our choice of couplings. 
We approximate the potential by the expression
\beq
V_\L(\phi_1, \phi_2) = 
V_\L(\phi_1) +\frac{1}{2} m_2^2(\L) \phi_2^2
+ \frac{1}{4} \lambda_2(\L) \phi_2^4 -
\frac{1}{2}\lambda_{12}(\L) \phi_1^2 \phi_2^2~.
\label{potphi1}
\eeq
The evolution of 
$m_2^2(\L)$, $\lambda_2(\L)$ and 
$\lambda_{12}(\L)$ is determined through the truncated eqs.
(\ref{trunc}). However, the full $\phi_1$ dependence is preserved through
the numerical integration of eq. (\ref{4dr}), with the eigenvalues of the
mass matrix ${\cal M}_\L^2$ given by 
\beq
\left[ {\cal M}_\L^2 \right]_1 = 
\frac{\partial^2 V_\L(\phi_1)}{\partial \phi^2_1}
~~~~{\rm and}~~~~~~
\left[ {\cal M}_\L^2 \right]_2 = 
m_2^2(\L) - \lambda_{12}(\L) \phi_1^2. 
\label{eigen}
\eeq
This treatment permits a reliable study of the order of the 
symmetry-breaking phase transition.
The appearance of secondary minima of the potential 
at some point in the evolution can be studied in detail. As a result,
we can distinguish between first-and second-order phase transitions. 
This is not possible when local expansions of the potential, such
as the one leading to eqs. (\ref{trunc}), are employed.

In fig.~2 we present the evolution of the potential for zero-temperature
parameters  
$m^2_1(\L_0)=m_2^2(\L_0)$, $\lambda_1(\L_0)=0.01$, $\lambda_2(\L_0)=0.3$,
$\lambda_{12}(\L_0)=0.05$. 
The location of this parameter choice on the plot of fig.~1 is 
denoted by a black square. It is within the region for which
inverse symmetry breaking is expected. 
The temperature is chosen very close to the critical one: 
$T_{cr}/m_1(\L_0) \simeq 33.3$. This value is in very good agreement 
with the result of ref.~\cite{Roos}, for the same choice of 
zero-temperature parameters. It deviates significantly from
the perturbative prediction $T_{cr}/m_1(\L_0) \simeq 24.5$.
We use the rescaled variables defined in eqs.~(\ref{resc}), which 
permit the identification of the fixed points that may be relevant
for the phase transition. 
We plot the derivative $\partial \Vb_\L / \partial \rhb_1$
of the potential
as a function of the variable $\rhb_1={\phb_1^2}/{2}$,
for decreasing $\L$.  
The early stages of the evolution of the potential, when its curvature
at the origin becomes negative, are not clearly 
visible in this plot. The reason is that 
$\partial \Vb_\L / \partial \rhb_1$ 
is very small for large $\L$. The important point in this 
figure is the flow of the potential towards a 
scale-invariant solution (marked by WF) during the 
later stages of the evolution. 
This solution corresponds to the Wilson-Fisher
fixed point of the effective three-dimensional theory. 
For $\L \rightarrow 0$ the theory leaves the fixed point 
and flows towards the phase with symmetry breaking.
The evolution can also be described in terms of 
the parameters defined in eqs. (\ref{param}). Their rescaled 
versions 
\beq
\bar{m}^2_{1,2}(\L)= \frac{m^2_{1,2}(\L)}{\L^2}
~~~~{\rm and}~~~~~~~
\bar{\lambda}_{1,2,12}(\L)= 
\frac{\lambda_{1,2,12}(\L) T}{\L}
\label{paramr}
\eeq
are obtained from the rescaled potential of eqs.~(\ref{resc})
through relations analogous to eqs.~(\ref{param}).
(Notice that $\bar{m}^2_{1}(\L)$
is given by the value of  $\partial \Vb_\L / \partial \rhb_1$ for $\rhb_1=0$
in fig.~2.)
We present the evolution of all these parameters in figs.~3 and 4.
In both these plots we observe that the mass term 
${m}^2_{1}(\L)$ turns negative at some scale $\L \sim T$. 
The mass term ${m}^2_{2}(\L)$ stays positive and grows as 
$\L$ decreases. For $\L \ll T$ we observe that
${m}^2_{2}(\L) \gg \L^2$ and 
the $\phi_2$ field decouples. 
Eventually, for $\L \rightarrow 0$, the renormalized parameters
${m}^2_{2}(\L)$ and $\lambda_2(\L)$ take constant values.  
After the decoupling of the $\phi_2$ field, 
only the fluctuations of the $\phi_1$ field contribute to the evolution. 
In this regime, the evolution equation for the potential can be 
written in the form of eq. (\ref{3drr}), with only the 
eigenvalue 
$\left[ \bar{\cal M}_{\L}^2 \right]_1 = 
\partial^2 \Vb_\L/ \partial \phb_1^2$ contributing. 
The scale-invariant solution of the resulting equation 
corresponds to the Wilson-Fisher fixed point of the 
one-scalar three-dimensional theory. 
Its explicit form can be seen in fig.~2. 
We conclude that the role of the $\phi_2$ field is to 
trigger the inverse symmetry breaking by inducing a 
negative mass term $m^2_1(\L)$ at the early stages of the
evolution. It subsequently decouples and the dynamics of the
phase transition is governed by the Wilson-Fisher fixed point of
the one-scalar three-dimensional theory. The resulting phase
transition is of second order. Its universal behaviour can be
parametrized in terms of critical exponents and amplitudes. 
For a detailed discussion see ref. \cite{Nick1}.

The evolution close to the fixed point is apparent in fig.~3, in the range
where the rescaled parameters 
$\bar{m}^2_1(\L)$, $\bar{\lambda}_1(\L)$ become 
constant.  
Their fixed-point values are
$\bar{m}^2_{1,fp} \simeq -0.44$, $\bar{\lambda}_{1,fp} \simeq 3.2$.
In this regime, the parameters 
${m}^2_1(\L)$, $\lambda_1(\L)$ evolve towards zero according to 
${m}^2_1(\L)=\bar{m}^2_{1,fp} \L^2$,
${\lambda}_1(\L)=\bar{\lambda}_{1,fp} \L/T$ (see fig.~4).
This explains why the fixed point is relevant very close to the
phase transition. If the temperature is chosen such that the
evolution stays close to the fixed point for a long `time'
$t=\log(\L/T)$, the curvature of the potential at the origin is
very small at the end of the evolution. Through sufficient fine-tuning
of the temperature, the curvature can be made arbitrarily small. 
This is the criterion for the occurrence of a second-order phase transition.
Notice that the renormalized quartic coupling $\lambda_1(\L=0)$ is also
expected to
be zero at the critical temperature \cite{Nick2}. 
According to eqs. (\ref{trunc}) 
the coupling $\lambda_{12}(\L)$ evolves as
$\lambda_{12} \sim \L^{a}$, with 
$a=3\bar{\lambda}_{1,fp}/2 \pi^2
(1+\bar{m}^2_{1,fp})^2 \simeq 1.6$.
For $\lambda_{2}(\L)$ we obtain
$\lambda_{2}(\L) = \lambda_{2}(\L_i) 
+c \left( \L^{2a-1}-\L_i^{2a-1} \right)/(2a-1)$, where
$\L_i$ is the scale at which the fixed-point solution is initially 
approached, and
$c$ a calculable constant.
The predicted evolution of $\lambda_{12}(\L)$ and $\lambda_{2}(\L)$  
is confirmed by fig.~4. If we had used the truncated equations 
(\ref{trunc})
for the discussion of the evolution of $m^2_1(\L)$
and $\lambda_1(\L)$, as in ref. \cite{Roos}, 
we would have obtained $a=1/3$. This would have predicted
an unphysical, singular behaviour 
$\lambda_{2}(\L) \sim - \L^{-1/3}$ for $\L \rightarrow 0$. 
Our solution of the 
evolution equation for the full potential along the $\phi_1$ axis 
resolves this problem, and
describes the decoupling of the $\phi_2$ field properly.

In conclusion, we have used the real-time formulation of the
exact renormalization group in order to study the question of 
inverse symmetry breaking in the context of the two-scalar theory
of eq. (\ref{pot}).  
We have verified the
conclusion of ref. \cite{Roos} that inverse symmetry breaking is 
confirmed by the renormalization-group approach. 
We have also determined the parameter range that leads to 
inverse symmetry breaking, through appropriate truncations
of the partial differential equation that describes the 
evolution of the potential. This parameter range is significantly
smaller than the perturbative prediction, in agreement with 
ref. \cite{Bimonte1}.
We have also obtained a numerical solution of the 
evolution equation for the full potential along the $\phi_1$ axis,
without relying on truncations in that direction. In this way we have
obtained a detailed picture of the symmetry-breaking phase transition.
The $\phi_2$ field triggers the symmetry breaking along the
$\phi_1$ direction and subsequently decouples. 
The phase transition is of second-order, governed by the Wilson-Fisher
fixed point of the effective three-dimensional theory.
Our improved treatment  gives no indication
of symmetry restoration 
for the range of temperatures above the critical one, in which we
obtain stable numerical solutions.

\noindent
{\bf Acknowledgements:}
We would like to thank M. D'Attanasio, B. Gavela, 
J. Orloff and G. Senjanovi\'c
for helpful discussions. M.P. has been supported by an E.C. Fellowship
under the H.C.M. program. 
N.R. has been supported in part by CICYT under grant AEN-96/1718
and by DGICYT under grant PB95-1077 (Spain).

\newpage

\section*{Figures}

\renewcommand{\labelenumi}{Fig. \arabic{enumi}}
\begin{enumerate}

\item  
The parameter space that leads to inverse symmetry breaking
($\lambda_2 = 0.3$).
\vspace{5mm}

\item  
The derivative $\Vb'_\L=\partial \Vb_\L / \partial \rhb_1$
of the potential 
as a function of $\rhb_1={\phb_1^2}/{2}$
for decreasing $\L$, for a theory very close to the phase transition.
\vspace{5mm}

\item  
The evolution of the rescaled masses and couplings for a theory very close to 
the phase transition.
\vspace{5mm}

\item  
The evolution of the masses and couplings for a theory very close to 
the phase transition.\vspace{5mm}

\end{enumerate}

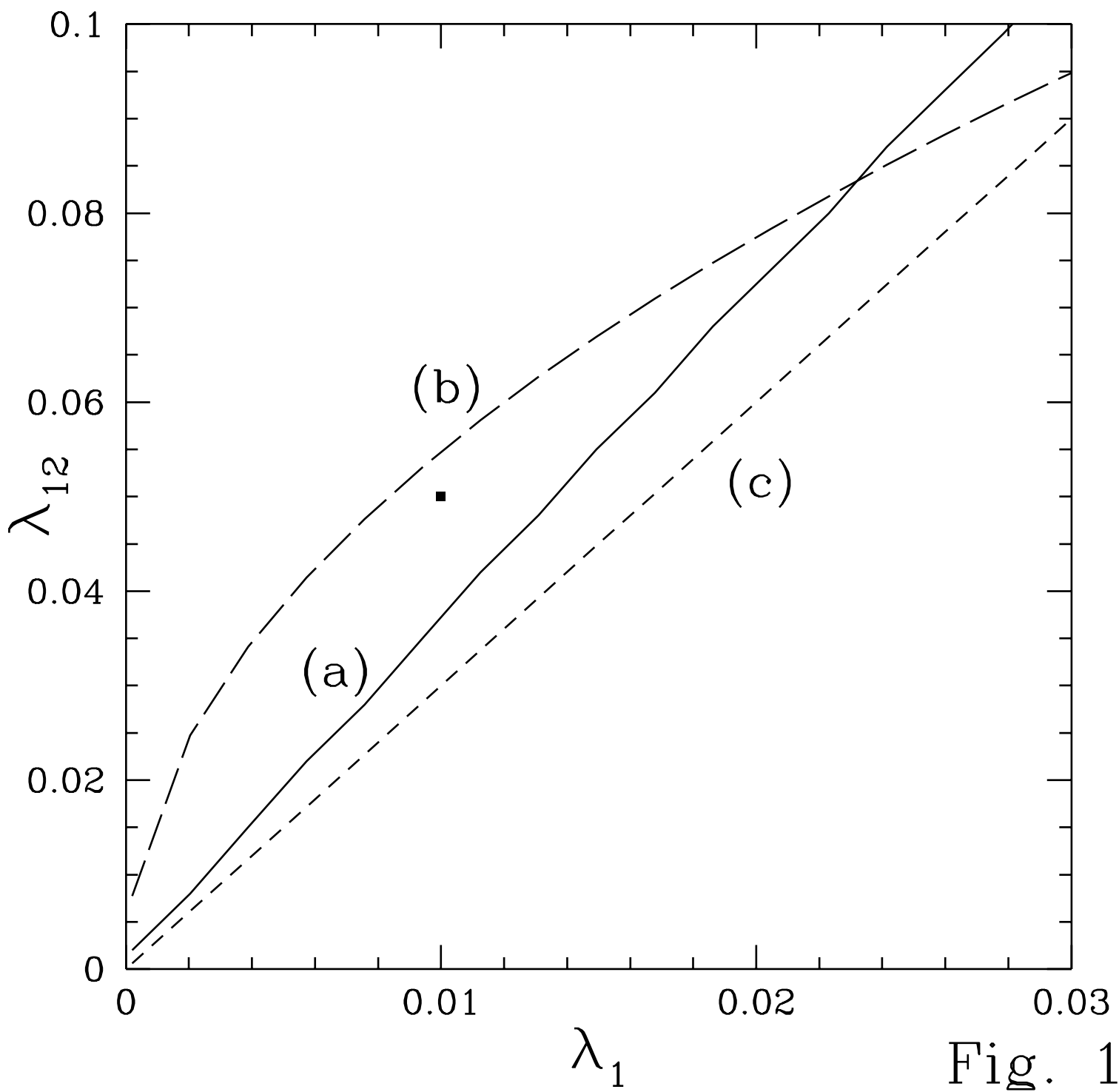
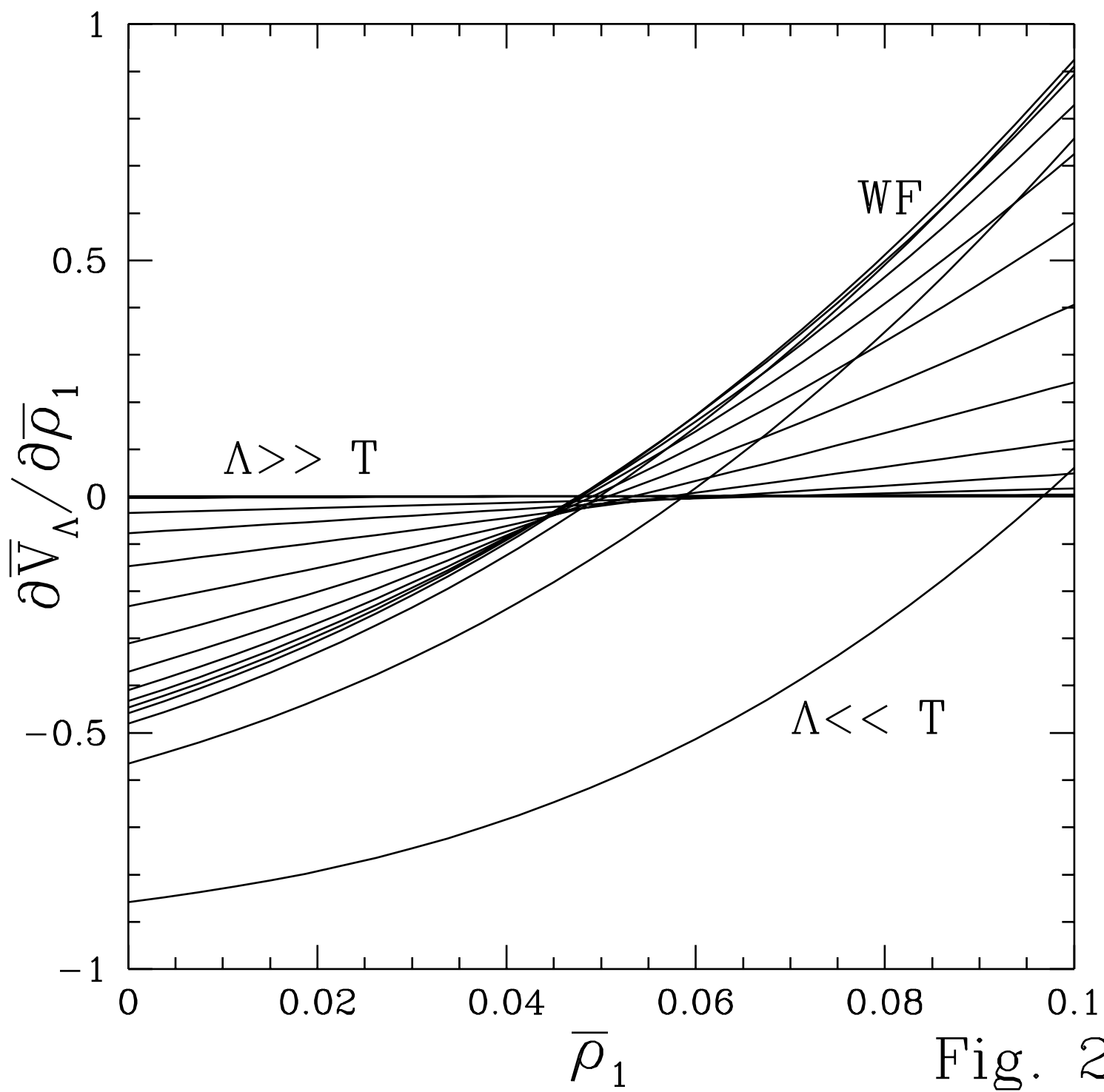
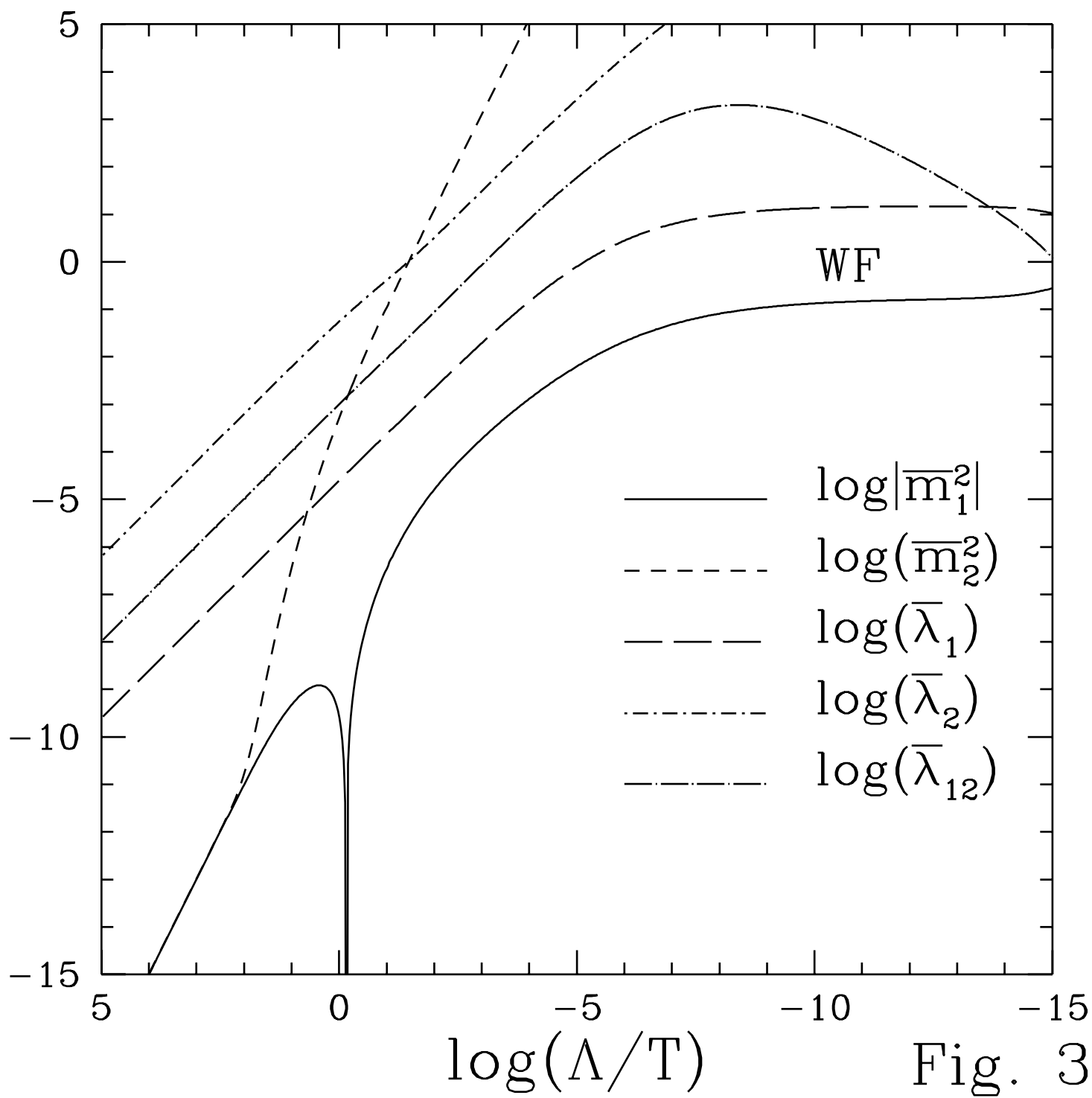
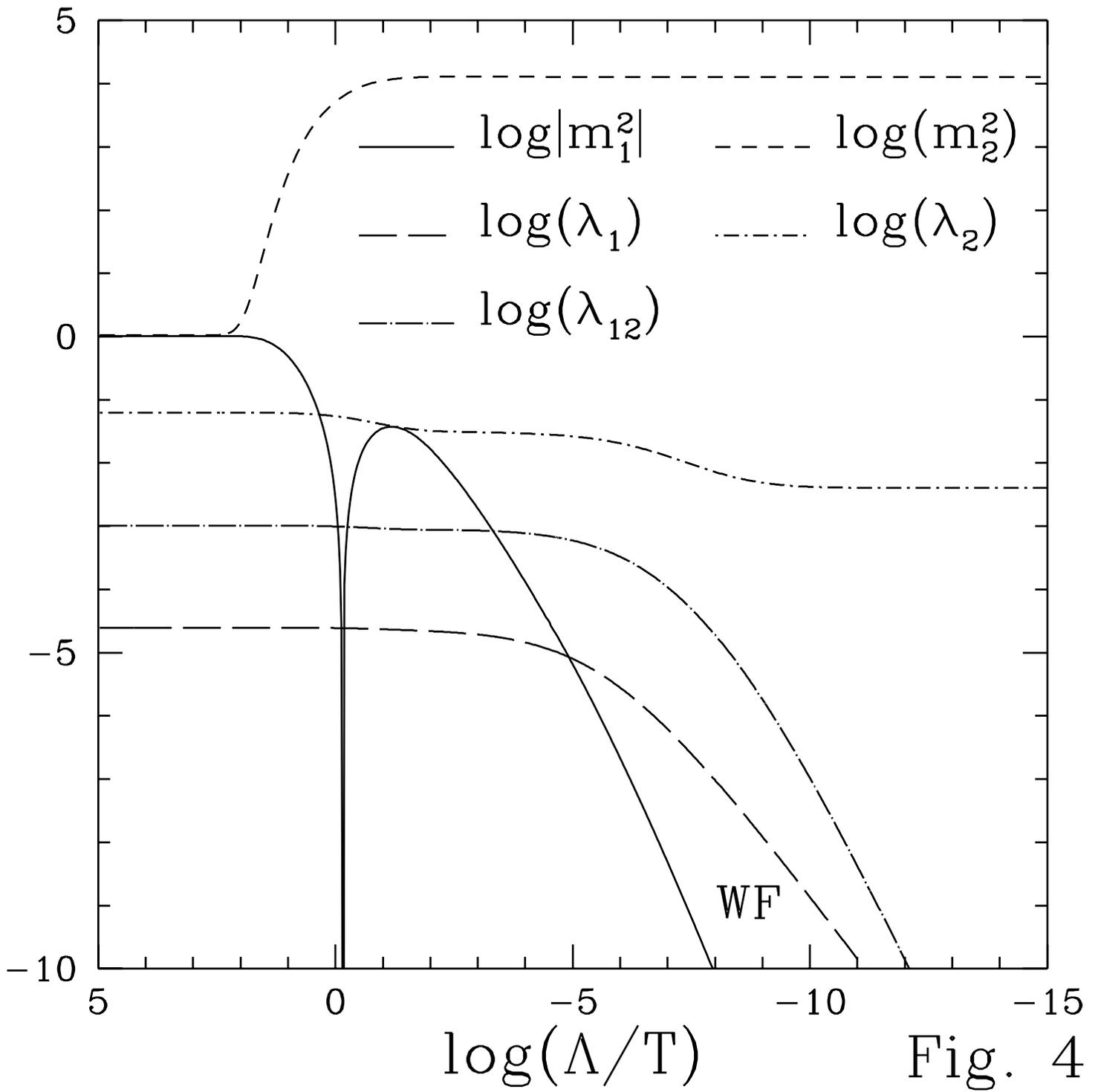
\end{document}